\documentclass[aps,pre,showkeys,showpacs,notitlepage,superscriptaddress,floatfix]{revtex4-1}
\pdfoutput=1
\usepackage{amsmath}
\usepackage{amssymb}
\usepackage{amsfonts}
\usepackage{graphicx}
\usepackage{hyperref}
\usepackage{color}
\usepackage{multirow}
\usepackage{mathtools}
\usepackage{booktabs}
\usepackage{amsthm}

\newcommand{\bel}[1]{\begin{equation}\label{#1}}

\newcommand{\be}{\begin{equation}}

\newcommand{\ba}{\begin{eqnarray}}
	\newcommand{\ea}{\end{eqnarray}}
\newcommand{\rf}[1]{(\ref{#1})}

\newcommand{\qe}{\end{equation}}

\hypersetup{pdftitle=Edge-based analysis of networks}

\begin{document}
\title{Edge-based analysis of networks: \\ Curvatures of graphs and hypergraphs}

\author{Marzieh Eidi}
\affiliation{Max Planck Institute for Mathematics in the Sciences, Leipzig 04103 Germany}
\author{Amirhossein Farzam}
\affiliation{Max Planck Institute for Mathematics in the Sciences, Leipzig 04103 Germany}
\author{Wilmer Leal}
\affiliation{Max Planck Institute for Mathematics in the Sciences, Leipzig 04103 Germany}
\affiliation{Bioinformatics Group, Department of Computer Science,
Universit{\"a}t Leipzig, 04107 Leipzig, Germany}
\author{Areejit Samal}
\affiliation{The Institute of Mathematical Sciences (IMSc), Homi Bhabha National Institute (HBNI), Chennai 600113 India}
\affiliation{Max Planck Institute for Mathematics in the Sciences, Leipzig 04103 Germany}
\author{J\"urgen Jost}
\email{jost@mis.mpg.de}
\affiliation{Max Planck Institute for Mathematics in the Sciences, Leipzig 04103 Germany}
\affiliation{The Santa Fe Institute, Santa Fe, New Mexico 87501 USA}

\begin{abstract}
	The relations, rather than the elements,  constitute the structure of networks. 
We therefore develop a systematic approach to the analysis of networks, modelled 
as graphs or hypergraphs, that is based on structural properties of (hyper)edges, 
instead of vertices. For that purpose, we utilize so-called network curvatures. 
These curvatures quantify the local structural properties of (hyper)edges, 
that is, how, and how well, they are connected to others. In the case of directed 
networks, they assess the input they receive and the output they produce, and 
relations between them. With those tools, we can investigate biological networks. 
As examples, we apply our methods here to protein-protein interaction, transcriptional 
regulatory and metabolic networks. 
\end{abstract}

\maketitle

\section{Introduction}
A central paradigm of structuralism \cite{DeSaussure16,Levi-Strauss58} is the analysis of structural relations regardless of the identity of the elements involved. That is, a structure is conceived in terms of the {\it relations} between elements. One wants to understand the types of relations, rather than the nature of the elements. This paradigm is obviously also fundamental for the analysis of empirical networks, be they from the biological sciences or other domains. Such an analysis then again abstracts from the specific content of the elements and concentrates on the formal relations between them. In that manner, one can both find universal features that hold across a wide range of networks from different domains, and properties that are specific to particular empirical domains. 

For that purpose, many different measures have been developed. Some of these measures, like for instance the assortativity (see for instance \cite{newman2003mixing,Jackson08}),  are of a global nature, that is, associate some number to the entire network. Such a number is usually an average or perhaps, like the diameter of a network, an extremum of locally measured quantities. In any case, the basis for such global measures is to first develop local measures. For a more refined analysis, one can then also look at the statistics of those local measures, instead of lumping them together in a single number (for instance \cite{piraveenan2010classifying,Farzam20a} for assortativity).

Some of these local measures require global computations in the network; for instance, for computing the diameter, one needs to evaluate the distances between any two elements. Therefore, some of these measures are difficult to evaluate in practice for networks of more than a moderate size. Others, including those that we shall concentrate on in this contribution, require only local computations and can be very easy to evaluate.

Now, somewhat surprisingly in view of the above structuralist paradigm, many of the local measures assign numbers to the elements of the network, rather than directly to its relations. The most basic one here is the degree of an element, the number of relations that it participates in. More global measures for instance evaluate the robustness of the network in terms of how many or which elements need to be eliminated in order to disconnect the network. See for instance \cite{Newman10,Estrada12}. 

In this situation, we and our collaborators have developed the research paradigm of a relation based analysis of networks (for instance \cite{Sreejith16a,Weber17a,Sreejith17a,Saucan18b,Samal18b,Saucan19b,Painter19,Saucan20a,Leal20b,Farzam20a}. That is, we evaluate relations and associate measures to them whose statistics across the network then can provide structural insight. 
 
There is another shortcoming of much of traditional network analysis. It tries to represent all structures as graphs, that is, considers only pairwise relations. For instance, a relation between three elements is simply broken up into three pairwise relations. That may, however, suppress some important structural insight. Take the example of scientific collaborations. From preprint repositories in the internet, it is easy to extract patterns of collaborations from coauthorships between authors. There are some single author papers, but of more interest are those written by several authors. There may be more than two authors involved in some paper, say $A,B$, and $C$. Of course, one could reduce it to pairwise relations and say that any two of them are coauthors. But there may be more structure. For instance, there may also exist a two-author paper by $A$ and $B$, no such paper between $A$ and $C$, and a paper of $B$ and $C$ with two other authors $D$ and $E$. This is obviously not captured by the pairwise relations, and for a more adequate model of the structure of scientific collaborations, we should rather use a hypergraph instead of a simple graph. In a hypergraph, a hyperedge can connect any number of elements. See for instance \cite{Berge85,Gallo93,Ghoshal09,Bretto13,joslyn2020hypernetwork}. In computer science, directed hypergraphs are also known as Petri nets \cite{Petri62,Petri08}. They were originally proposed by Petri as models of chemical reactions. Over the years, while not as widely employed as graphs, they have found applications in many fields, for instance recently as models of coupled dynamics in statistical physics  \cite{Mulas20a,Banerjee20,Battiston20},   
 of social contagion
\cite{de2020social}  and for knowledge representation in natural language processing \cite{menezes2019semantic}.   

In this contribution, we shall summarize relation-based measures both for graphs, that being the simplest case, and for hypergraphs.

\section{The idea of curvatures}
The name {\it curvature} derives from its origin in differential geometry. Originally, curvature was an infinitesimal quantity, obtained by taking second derivatives of functions describing shapes of smooth objects, like curves or surfaces. In Riemannian geometry, curvatures obtained a deeper conceptual significance, as tensors encoding the geometric invariants of Riemannian metrics of  smooth manifolds \cite{Jost17a}. In particular, the Ricci tensor is fundamental not only in Einstein's theory of general relativity and in elementary particle physics (the Calabi-Yau manifolds of string theory, for instance, are characterized by the vanishing of the Ricci tensor), but it also permeates much of modern research in Riemannian geometry. While Ricci curvature in Riemannian geometry again is computed infinitesimally, by taking second derivatives of the metric tensor, it essentially encodes local property, like the average divergence of geodesics or the growth of the volume of balls as a function of their radii. Moreover, Bochner type identities link it to other important geometric quantities, like the first eigenvalue of the Laplace operator. See for instance \cite{Bauer17} for a survey. 

Since such objects and properties are also meaningful and important in metric spaces that are more general than Riemannian manifolds, alternative definitions of Ricci curvatures have been proposed that are formulated in terms of local quantities and no longer depend on taking derivatives. Several of these definitions turned to be also meaning- and useful for graphs, and we have extended them to hypergraphs and are exploring their properties. Here, we shall not recount the history in detail, but rather systematically develop a conceptual approach that is in line with the paradigm of structuralism described at the beginning. We only note the curious fact that these concepts, although extremely natural from a structuralist perspective, were not developed directly, but inspired by concepts in a different, and more highly developed branch of mathematics, Riemannian geometry. 

\section{How relations connect}
Abstractly, there are different types of relations. They can vary with respect to the number of elements involved, they can be symmetric or directed, that is, distinguish between inputs and outputs, and they may also carry weights. The simplest case are binary, symmetric and unweighted relations. Such a web of relations is then modelled by an undirected and unweighted graph whose vertices stand for the elements in question and whose edges represent the presence of a relation between the two vertices they connect. For simplicity, we also assume that the graph is simple, that is, there is at most one edge between any two vertices, and that it is connected, that is, by passing from edge to edge we can reach any vertex from any other one, although these assumptions are not essential for any of the sequel. So, we start with that case.

We want to assess how a relation, that is, an edge of such a graph, sits in the web of relations, that is, how it relates to other relations. Two edges are called neighbors when they share a vertex. We can then already define the simplest concept, called {\it Forman-Ricci curvature}, because it was introduced by Forman \cite{Forman03} as an analogy with the Ricci curvature of Riemannian geometry (the analogy relates to the role it plays in Bochner type identities). We define the degree of an edge $e$ as
\bel{1}
\deg (e):= \# (\mathrm{neighbors\ of }\ e),
\qe
and define its Forman-Ricci curvature as
\bel{2}
F(e):=2-\deg (e).
\qe
The $2$ and the minus sign are somewhat unfortunate for our purposes, but they are there because of the analogy with the well-established  Ricci curvature of Riemannian geometry, and they are useful from an abstract geometric perspective.

When the edge $e$ connects the vertices $v,w$, we can also assess their contribution to the number of neighbors of $e$. We let $\deg_v(e)$ be the number of edges that share with $e$ the vertex $v$.
Then, obviously,
\bel{3}
F(e)=2 -(\deg_v (e) +\deg_w (e)).
\qe
Instead of the sum of the degrees, we may also consider their difference. When the edge is not directed, there is no intrinsic structural difference between the two vertices that it connects, and so, it is natural to take the absolute value of the difference and define the {\it degree difference} \cite{Farzam20a} as
\bel{4}
\daleth(e):=|\deg_v (e) -\deg_w (e)|.
\qe
Let us interpret the geometric significance of these quantities. $\daleth(e)$ is large when $e$ connects vertices of different types, a well-connected one from which many further edges emanate, and a less well-connected one from which only fewer edges originate. The statistics of this quantity therefore quantify to what extent the network is assortative, that is, typically connect similar vertices (small $\daleth(e)$), or disassortative, that is, typically connect dissimilar vertices (large $\daleth(e)$). This is important, for instance, because social networks tend to be assortative 
\cite{fisher2017perceived} (well connected people like to link with other well connected people, and this further improves their position in social networks). In contrast, $F(e)$ is very negative, that is, has a particularly large absolute value when both ends of an edge are well connected. Such edges may play a very important role in the network. In fact, we have found \cite{Samal18b} that a quantity that needs a global computation, edge-betweenness centrality (see \cite{Newman10}), is statistically well correlated with $F(e)$. This edge-betweenness centrality measures how many shortest connections between  pairs of vertices in the network pass through that particular edge. The computation of that quantity is expensive because all shortest connections between any two vertices have to be evaluated. In contrast, the computation of $F(e)$ is very quick and easy, because only local neighborhoods have to be evaluated. 

Edges with large $|F(e)|$ also play an important role for spreading in the
network because from its vertices many other vertices in the network can be
reached in a single step. There is one issue here, however. Edges from the two
vertices $v,w$ of $e$ may end at the same vertex $z$, that is, $v,w,z$ may
form a triangle. In that case, they would not  contribute to spreading into
different directions. Or the endpoint of an edge from $v$ and that of an edge
from $w$ may be connected themselves by an edge. That is, they form a
quadrangle together with $v$ and $w$. Again, that does not really constitute
spreading into different directions. It is possible to address this issue by
inserting two-dimensional faces into such triangles and perhaps also into
quadrangles, and then to evaluate the Forman curvature of the resulting
simplicial or polyhedral complex. Those faces would then increase the Forman
curvature and make it less negative or even positive. See for instance
\cite{Saucan19b}. 

This aspect is  taken care of in a different way by a more refined concept of Ricci curvature, the {\it Ollivier-Ricci} curvature introduced in \cite{Ollivier07}. For that purpose, consider the edge $e=(v,w)$ and let $e_v=(v,v_1)$ and $e_w=(w,w_1)$ be edges emanating from $v$ and $w$, respectively. We then define their distance w.r.t. $e$ as
\bel{5}
d_e(e_v,e_w):= d(v_1,w_1)
\qe
where $d(v_1,w_1)$ denotes the distance between $v_1$ and $w_1$ in the network, that is, the minimal edges that have to be traversed for getting from $v_1$ to $w_1$. Let $E_v$ be the set of edges that have $v$ as a vertex, and let $|E_v|$ be its cardinality. We then define a probability measure $\mu_v$ on the set of all edges $E$ by giving each edge $e_v\in E_v$ the weight $\frac{1}{|E_v|}$ and all edges not in $E_v$ the weight 0. We then define the Ollivier-Ricci curvature \cite{Ollivier07} of the edge $e=(v,w)$ as
\bel{6}
O(e):=1-W_1(\mu_v,\mu_w)
\end{equation}
where $W_1$ is the 1-Wasserstein distance between $\mu_v$ and $\mu_w$, 
\begin{equation}\label{7}
W_1(\mu_v,\mu_w):=\inf_{p\in \Pi(\mu_v,\mu_w)}\sum_{(e_1,e_2)\in E\times E} d_e(e_1,e_2) p(e_1,e_2)
\end{equation}
and $\Pi(\mu_v,\mu_w)$ is the set of measures on $E\times E$ that  project to $\mu_v$ and $\mu_w$, resp.
We thus try to arrange the two collections $E_v,E_w$ of edges sharing one of their endpoints with $e$ in an optimal manner, that is, that the average distances of the arranged pairs become as small as possible. We note that the sets $E_v$ and $E_w$ both include the edge $e=(v,w)$ that we are evaluating. This convention is only needed to let our definition agree with that originally proposed in the literature, but could otherwise be abandoned, to make the definition more natural in the present context.

In order to evaluate \rf{6}, we have to optimize the arrangement between the edges in $E_v$ and $E_w$, to make the transportation cost as small as possible. Since this is a quantity all edges in those two edge sets, it is not necessarily the case that an optimal transport plan arranges each edge $e_1$ in $E_v$ with the edge $e_0$ in $E_w$ closest to it. There might be some competition, as there might be other edges $e_2,e_3,\dots $ for which $e_0$ is closest. But even if there is no such competition, it might be overall more beneficial to arrange $e_1$ with an edge different from $e_0$. Also, because of the normalization, the edges in $E_v$ and $E_w$ have fractional weights, and if the cardinalities of the two edge sets are different, also the corresponding weights are different, necessitating an arrangement where some part of an edge in $E_v$ is arranged with some part of an edge in $E_w$, and other parts with other ones. 

Notwithstanding these complications, let $m_i$ be the fraction of edges in $E_v$ that are moved a distance $i$ in some optimal transport plan (such an optimal arrangement  need not be unique, but that does not matter for our discussion). Then \cite{Eidi2020}
\bel{8}
O(e)=m_0-m_2-2m_3.
\qe
In particular, moving an edge a distance 1 does not contribute at all to $O(e)$. (While $m_1$ itself does not appear in \rf{8}, its computation is nevertheless needed as an intermediate step for computing $m_2$ and $m_3$.) Distance 0, that is, when $e$ participates in a triangle, has a positive contribution. A pentagon, that is, distance 2, has a negative contribution, but not as a negative as the maximal distance, that can occur in a transportation plan, which is 3. This simple formula thus encodes the essential features of Ollivier-Ricci curvature. In fact, we could simply take \rf{8} as the definition of $O(e)$, instead of utilizing the more complicated formula \rf{7}.

More generally, the Ollivier-Ricci curvature is related to the clustering coefficient, that is, the relative frequency of triangles in the network \cite{Jost14}.

\subsection*{Protein-protein interaction networks}
\begin{figure}[t]
	\centering
	
	\includegraphics[width=.75\textwidth]{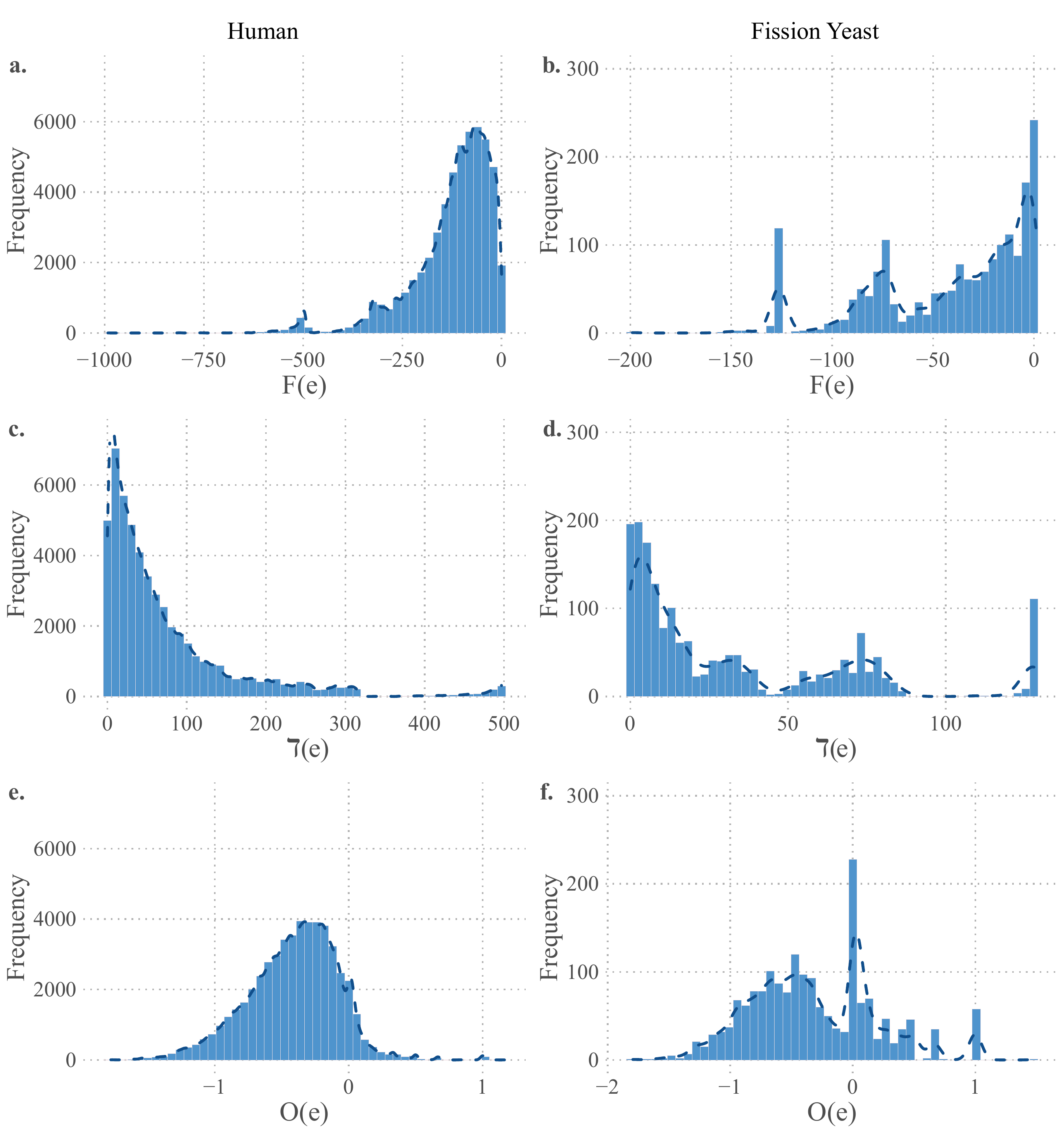}
	
	\caption{The distribution of \textbf{(a,b)} Forman-Ricci curvature, 
		\textbf{(c,d)} degree difference, and \textbf{(e,f)} 
		Ollivier-Ricci curvature in the giant components of the 
		binary protein interaction networks in human (left) 
		and fission yeast (right), respectively. In each case,
		protein-protein interactions are represented via an undirected 
		and unweighted graph. The nodes and edges represent proteins 
		and binary interactions between them, respectively. 
		The giant component of the human network has 8152 nodes and 
		52036 edges, while of the fission yeast network has 1306 nodes 
		and 2278 edges.}
	\label{fig:ppis}	
\end{figure}

To illustrate an application of these structural measures to empirical 
data, we have studied the protein-protein interaction (PPI) networks in human 
\cite{luck2020reference}, with 8275 nodes and 52569 edges, and fission 
yeast \textit{S. pombe} \cite{vo2016proteome}, with 1306 nodes and 2278 edges.
The edges in these network represent binary interactions between the pair of proteins represented as nodes.
These undirected and unweighted networks are disconnected with several 
components, however, they both include a giant component. The giant component 
consists of 8152 nodes and 52036 edges in the human PPI network, and of 
1306 nodes and 2278 edges in fission yeast PPI network. We have computed the 
Forman-Ricci curvature, Ollivier-Ricci curvature and degree difference of 
edges in these networks, and their distributions are shown in 
Figure \ref{fig:ppis}.

In the human PPI network, while Ollivier-Ricci curvature has a unimodal 
distribution, the bimodal distribution of Forman-Ricci curvature in 
Figure \ref{fig:ppis} signals an evident heterogeneity in the space of 
protein-protein interactions in the giant component; a major group of interactions 
are distributed around a relatively small-valued mode, and a small group of 
interactions between proteins that are, in average, involved in a signficantly 
larger number of interactions. The degree difference distribution indicates that, 
although the majority of interactions are between proteins with relatively similar 
degree, a noticeable proportion of the edges have a considerably large degree 
difference, which can be as large as $497$. This observation is in line with the 
fact that this network is moderately disassortative with assortativity value 
$\sim -0.119$.

Unlike the Ollivier-Ricci curvature distribution of the human PPI network, 
the fission yeast PPI network has a trimodal distribution of the Ollivier-Ricci 
curvature, reaching its global mode at curvature value $0$. In fact, in the PPI 
network for fission yeast, all three measures have multimodal distributions, as 
demonstrated in Figure \ref{fig:ppis}. Interestingly, the peaks over highly 
negative values of Forman-Ricci curvature have larger frequencies than those over 
the moderately negative values. A similar phenomenon is observed in the degree 
difference distribution of the fission yeast PPI network. The global degree 
assortativity of the fission yeast PPI network is $\sim -0.237$. This means that 
the fission yeast PPI network is considerably more disassortative than the human 
one, which is explained by the more substantial proportion of interactions in 
fission yeast between proteins with significantly different degrees. 
Thus, we see that the distribution of curvature and degree difference 
values can point us to biologically relevant properties of the interaction 
statistics in the PPI networks of different species.

\section{Directed graphs}
It is not only the case that the preceding constructions extend to directed
graphs, but in fact, they become even more natural in that context. Curvature
concepts for directed graphs have been systematically developed and evaluated
in \cite{Saucan19b} (see also see \cite{Saucan18b}). Here,  we shall formulate the concepts in such a manner
that they will naturally generalize to hypergraphs.

Thus, let  $e=[v,w]$ be a directed edge with tail $v$ and head $w$, that is, going from $v$ to $w$. The input of $e$ at its tail $v$ then are all edges that have $v$ as their head; let their number be $\deg_{in} (e)$. Similarly, $\deg_{out} (e)$ denotes the number of output edges of $e$, that is, those that have $w$ as their tail. We may then put \cite{Leal18a}
\bel{11}
F_{\rightarrow{}}(e):= 2 - \deg_{in} (e) -\deg_{out} (e).
\qe
We could also form alternative expressions by considering the numbers of edges that have $v$ as their tail and/or of those that have $w$ as their head. Similarly for the next expression, the directed degree difference \cite{Farzam20a} 
\bel{12}
\daleth_{\rightarrow{}}(e):= \deg_{out} (e) - \deg_{in} (e) .
\qe
$F_{\rightarrow{}}(e)$ now is very negative, or equivalently, $\deg_{in} (e) +\deg_{out} (e)$ is very large for those edges that receive a lot of input and produce a lot of output. $\daleth_{\rightarrow{}}(e)$ is positive for those edges that are productive in the sense that they produce more output than they receive input, or that lead to more diversification. It is negative for edges that are receptive, that is collect more input than emit as output. 

Likewise, we can define the Ollivier-Ricci curvature $O_{\rightarrow{}}(e)$ of a directed edge \cite{Eidi2020} by computing the optimal transportation distance between its input and its output. When there are no shorter connections between inputs and outputs than those going through $e$ itself, then $O_{\rightarrow{}}(e)$ assumes its smallest possible value $-2$. In contrast, when inputs coincide with outputs, that is, if there is a directed triangle from a vertex $u$ to itself, where $u$ produces both an input of $e$ and receives an output of $e$, then this yields a positive contribution. In fact, formula \rf{8} perfectly extends to the directed case.  

Let us recall the procedure in detail. For the directed edge $e[v,w]$, we define two measures,
\ba
\label{13} \begin{cases}
\text{if }e\text{ has no incoming edges: }& \mu_{in}(e)=1\\
\text{if }e\text{ has }n_1\text{  incoming edges: }& \mu_{in}(e_1)=\frac{1}{n_1} \text{ for each incoming edge}
\end{cases}\\
\label{14} \begin{cases}
\text{if }e\text{ has no outgoing edges: }& \mu_{out}(e)=1\\
\text{if }e\text{ has }n_2\text{  outgoing edges: }& \mu_{out}(e_2)=\frac{1}{n_2} \text{ for each outgoing edge}
\end{cases}\\
\nonumber
\text{and } \mu_{in}(e')=\mu_{out}(e')=0 \text{ for all edges }e' \text{ not occurring in those formulae}.
\ea
The first cases, that is, where there are no incoming  edges at the tail or no outgoing edges at the head, that is, where the tail is a source or the head is a sink, represent complications that would not arise in the undirected case. As they are easily handled, we shall mostly ignore them. In any case, 
 both measures are normalized to have total mass 1 and thus are probability measures. 
We then define the distance between an edge $e_1$ occurring in \rf{13} and an edge $e_2$ occuring in \rf{14} as
\ba
\nonumber
d_e(e_1,e_2)=&\text{minimal number of edges needed to get from the tail of }e_1\\
&\text{ to the head of }e_2.
\label{15}
\ea
We then put again 
\bel{16}
O_{\rightarrow{}}(e)_{\rightarrow{}}:=1-W_1(\mu_{in},\mu_{out})
\end{equation}
where $W_1$ now is the 1-Wasserstein distance between $\mu_{in}$ and $\mu_{out}$, 
\begin{equation}\label{17}
W_1(\mu_{in},\mu_{out}):=\inf_{p\in \Pi(\mu_{in},\mu_{out})}\sum_{(e_1,e_2)\in E\times E} d_e(e_1,e_2) p(e_1,e_2)
\end{equation}
and $\Pi(\mu_{in},\mu_{out})$ is the set of measures on $E\times E$ that  project to $\mu_{in}$ and $\mu_{out}$, respectively. $E$ here is the set of directed edges of the directed graph under consideration. 
We again have the important formula \cite{Eidi2020} 
\bel{18}
O_{\rightarrow{}}(e)=m_0-m_2-2m_3,
\qe
where $m_i$ is the number of edges that have to be transported by the distance
$i$ in an optimal transport plan. Again, for two given edges in the in- and
output of $e$,  that distance might be larger than the distance \rf{15}.\\
Thus, the directed curvature notions evaluate flows through edges. As in
\cite{Saucan19b}, we may
also evaluate flows through vertices by taking the difference between the sum
of the Ricci curvatures of the incoming edges and that for the outgoing
edges. Moreover, in \cite{Saucan19b}, also notions of augmented Forman curvature were
developed for directed networks. Augmentation means that one inserts
two-dimensional faces into triangles of edges. Such triangles then increase
the curvature, and thereby decrease the difference between Forman and Ollivier
type curvatures. Here, however, we do not explore that direction.

\subsection*{Transcriptional regulatory networks}

\begin{figure}[t]
\centering
\includegraphics[width=.8\textwidth]{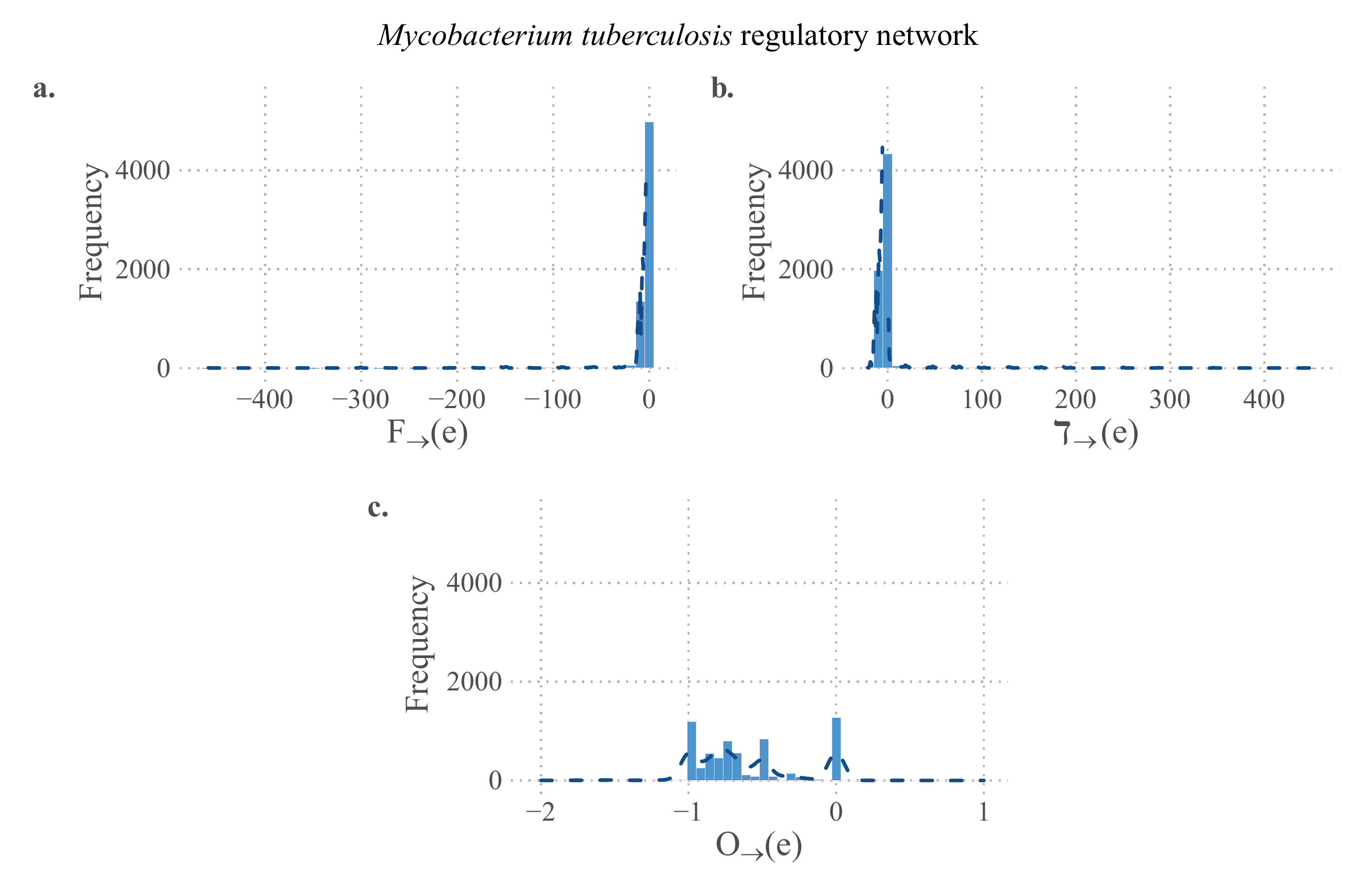}
\caption{The distribution of \textbf{(a)} Forman-Ricci curvature, 
\textbf{(b)} degree difference, and \textbf{(c)} 
Ollivier-Ricci curvature in the transcriptional regulatory network 
of \textit{Mycobacterium tuberculosis}. There are 6581 unweighted 
directed edges and 2547 unweighted nodes. The source in each directed 
edge is a transcription factor (TF) and the target is a target gene 
controled by the TF. 
}
\label{fig:mtb}	
\end{figure}

To illustrate an application to directed networks, we have studied the
transcriptional regulatory network (TRN) of the important human pathogen
\textit{Mycobacterium tuberculosis} \cite{Minch2015}, with 2547 nodes
and 6581 edges. The \textit{M. tuberculosis} TRN was constructed based
on ChIP-seq data for 143 transcription factors (TFs) \cite{Minch2015}.  
In this directed and unweighted network, each directed edge signifies the
regulatory control by a TF of a target gene. In other words, the source
nodes in this directed network are TFs while target nodes are target genes.
In Figure \ref{fig:mtb}, we show the distribution of the Forman-Ricci curvature, Ollivier-Ricci curvature and degree difference of directed edges in the \textit{M. tuberculosis} TRN. 
In Figure \ref{fig:mtb}a, it is seen that the edges are densely concentrated around Forman-Ricci curvature value $0$, and this indicates that the majority of the edges have a tail vertex with small in-degree and a head vertex with small out-degree. 
Likewise, most edges have zero or small value of directed degree difference, and this indicates that the in-degree of the tail vertex and the out-degree of the head vertex for most edges are rather similar. 
There are also $24$ vertices with out-degree greater than $100$, which can explain the long tail in both Forman-Ricci curvature and degree difference distributions in Figure \ref{fig:mtb}. 
On the other hand, the Ollivier-Ricci curvature of the edges in this TRN has a multimodal distribution, with major peaks corresponding to curvature values $0$, $-1$, $-0.5$, and $-0.75$.

\section{Weighted graphs}
The extension of all discussed concepts to weighted graphs is straightforward. One simply counts each edge with its weight. It is therefore not necessary to develop the details here.

\section{Hypergraphs}
The preceding concepts are set up in such a manner that they naturally extend to hypergraphs. We directly consider directed hypergraphs. A directed hypergraph may have several nodes through which it receives inputs and several nodes through which it produces outputs. An important example is a chemical reaction whose input nodes are called educts and whose output nodes products. In a chemical reaction, there may exist catalyzers, that is, substances that increase the rate of a reaction without being modified by it. Formally, they should be counted as both input and output nodes. That is, the two subsets of the nodes of a directed hypergraph, its input and output nodes, need not be disjoint. That will not constitute a problem for the formal concepts to be developed (see, for instance, \cite{Leal20b} for a discussion of directed hyperloops and their curvature).

A directed hypergraph $H=(V,E)$ consists of  a set  $V$ of nodes or vertices and  a set $E$ of ordered pairs of subsets  of $V$, not both of them being empty, called hyperedges.  For a hyperedge $e=(e_1,e_2) \in E$, $e_1\subset V$ is the  head of $e$, and $e_2\subset V$ is its tail. We let $|f|$ be the number of vertices in $f\subset V$. We let $\deg_{in}(e)$ of a hyperedge be the number of hyperedges that have an input node of $e$ as their head, and $\deg_{out}(e)$ the number of hyperedges that have an output node of $e$ as their tail. Since an input edge might connect to more than one input node of $e$, input hyperedges are counted with the number of input nodes of $e$ that they connect to, and analogously for output edges.  As in \rf{11}, we then define the Forman-Ricci curvature of a hyperedge $e=(e_1,e_2)$ as \cite{Leal18a}

\begin{equation}
   F_{\rightarrow{}}(e)=|e_1|+|e_2|-\deg_{in}(e)-\deg_{out}(e).
    \label{21}
\end{equation}
(For a different definition of the Forman-Ricci curvature of a directed hypergraph, see \cite{Saucan18a}.) 
Thus, here we count the number of inputs received through input nodes and the number of outputs produced at output nodes. As in \cite{Leal18a}, one can also define different types of Forman-Ricci curvature of a directed hyperedge by arranging inputs and outputs differently. 

Similarly, as in \rf{12}, we can put
\bel{22}
\daleth_{\rightarrow{}}(e):= \deg_{out} (e) - \deg_{in} (e) .
\qe

Following \cite{Eidi2020}, we can also define the Ollivier-Ricci curvature of a directed hyperedge via  the Wasserstein distance between two probability measures associated to a directed hyperedge. As in \rf{13} and \rf{14}, we need to define the corresponding measures $\mu_{in}, \mu_{out}$. In \rf{13} and \rf{14}, the principle that was that the total measure $1$ is evenly split among the inputs or the outputs, resp., unless we had a source or a sink. Now, there are more demands for splitting. A directed hyperedge may in general have more than one tail or head node, and at each of them, several incoming resp. outgoing hyperedges might be found, and each them may again have more than one tail or head. The principle then is to split the available measure at each step evenly among all the possible recipients.  We shall explain the resulting splitting procedure for $\mu_{in}$ as the one for $ \mu_{out}$ is analogous. Let the tail $e_1$ of the hyperedge $e=(e_1,e_2)$ have $\eta_{in}$ elements. A source, that is, an element  of $e_1$ without incoming hyperedges, gets the weight $\mu_{in}(v)=\frac{1}{\eta_{in}}$. To handle the others, we  define the set $\mathcal{M}$ of masses of $e=(e_1,e_2)$ as the union of the tails of hyperedges that come in at an element $w\in e_1$, that is, have $w$ in their head set. We then first divide the measure $\frac{1}{\eta_{in}}$ that we can distribute at such a $w$ evenly among all those incoming hyperedges, and for each such hyperedge, we divide the measure associated with it in that manner evenly in its tail. In that manner, we assign a measure to every element in $\mathcal{M}$. Thus, we have distributed the total measure 1 among the sources and the masses of our hyperedge. This yields $\mu_{in}$, and as mentioned,  $ \mu_{out}$ is constructed analogously by assigning measures to the sinks, that is, those members of $e_2$  and the holes, that is, the heads of hyperedges that have an element of $e_2$ in their tail set. 
 The (directed) distance $d(u,v)$ between  a mass $u$ and a hole $v$ of a hyperedge $e=(e_1,e_2)$ is defined as the minimal number of directed hyperedges connecting them. Again, it  is at most 3, and this value is attained if $u\to e_1$, $e_2\to v$ and there is no shorter way to move from $u$ to $v$ than to go through $e$. It is $0$ when $u=v$ is at the same time a mass and a hole of $e$, and it is $1$ if $u$ is an input of a hyperedge and $v$ is an output. Again, formally, we want to solve an optimal transport problem for  moving  the first probability measure to the second one. We thus  minimize
\begin{equation}
    \sum_{u\to e_i}\sum_{e_j\to v} d(u,v)\mathcal{E}(u,v) 
    \label{23}
\end{equation}
 over the set of  all matrices $\mathcal{E}$ (transport plans) whose entries $\mathcal{E}(u,v)$ represent the amount of mass from  $\mu_{\mathcal{M}}(u)$ moved from vertex $u$ to vertex $v$. 

\begin{figure}[th!]
	\centering
\includegraphics[width=7 cm, height= 5 cm]{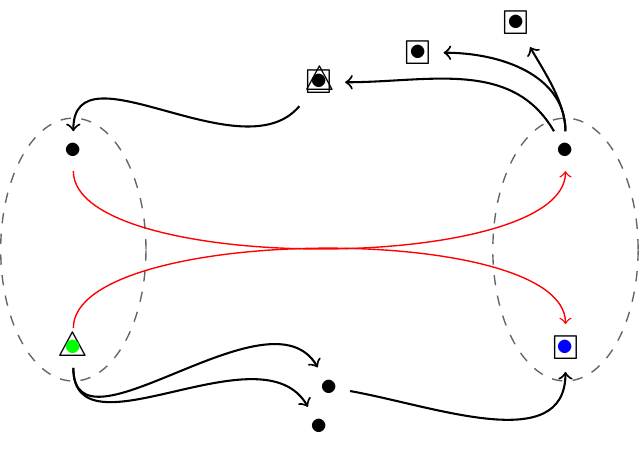}

 \caption{The red hyperedge is negatively curved as in an optimal transference
   plan, the size of coincident masses (triangles) and holes (quadrangles),
   located on the top middle vertex, is less than the size of the masses which
   need to be moved with distance 2. Also the two colored vertices in the left
   and the right of the figure are a source and a sink since they have no incoming resp. outgoing hyperedges.} 
   
  \label{fig:hyperedge}
 \end{figure}

If $m_\delta$  is the amount of  mass that is moved at distance $\delta$ in an optimal transport plan,  the directed Ollivier-Ricci curvature  of $e$ is defined as in \rf{14} and  becomes again as in \rf{15} 
\begin{equation}\label{24}
    O_{\rightarrow{}}(e)=m_0-m_2-2m_3.
\end{equation}
It is bounded above by $1$. This is reached when $m_0=1$, i.e., when each mass coincides with a hole of the same size. It is bounded   below by $-2$, reached  when $m_3=1$, i.e., when there are no shortcuts available and  each mass has to be moved through $e$ to reach a hole. Again, \rf{24} can be taken as the definition of $O_{\rightarrow{}}(e)$. While it depends on identifying an optimal transport plan, the formula as such is obviously very simple. For applications, see \cite{Leal20a}. 
 \\
For instance, we can consider the red hyperedge in Figure \ref{fig:hyperedge}.
Bullets represent vertices. The green bullet in the left is a source since it has no incoming hyperedge
 while the blue bullet in the right is a sink since it has not outgoing hyperedges. For representing masses and holes we use triangles and squares respectively.
As  the red hyperedge has two vertices in its tail set and each of them has at most one vertex as an incoming neighbour, the size of the masses is 1/2. 
In contrast, the sizes of the four holes are different. 
 The biggest one is  the sink, with mass 1/2. 
Another hole with the size  1/4 is  the top middle vertex which already 
got 1/2 of the total mass. The size of the remaining hole is 1/4, divided equally 
among the two vertices in the top right of the figure. Thus, both the triangles and the squares have total size 1, and the task now is to move the 
 triangles to the squares with least  total cost.  There are 
two optimal plans, leading to   the  negative curvature value -1/4;
 In one plan, $ m_0 = 1/4 $, $m_1 = 1/2$, $m_2 = 0$ and $m_3 = 1/4$, while in the other 
 $ m_0 = 1/4 $, $m_1 = 1/4$, $m_2 = 1/2$ and $m_3 = 0$.
 Note that there are also transfer plans with $m_2 = 1 $, and all other values 0,  but they are not optimal.

\begin{figure}[th!]
	\centering
	\includegraphics[width=.5\textwidth]{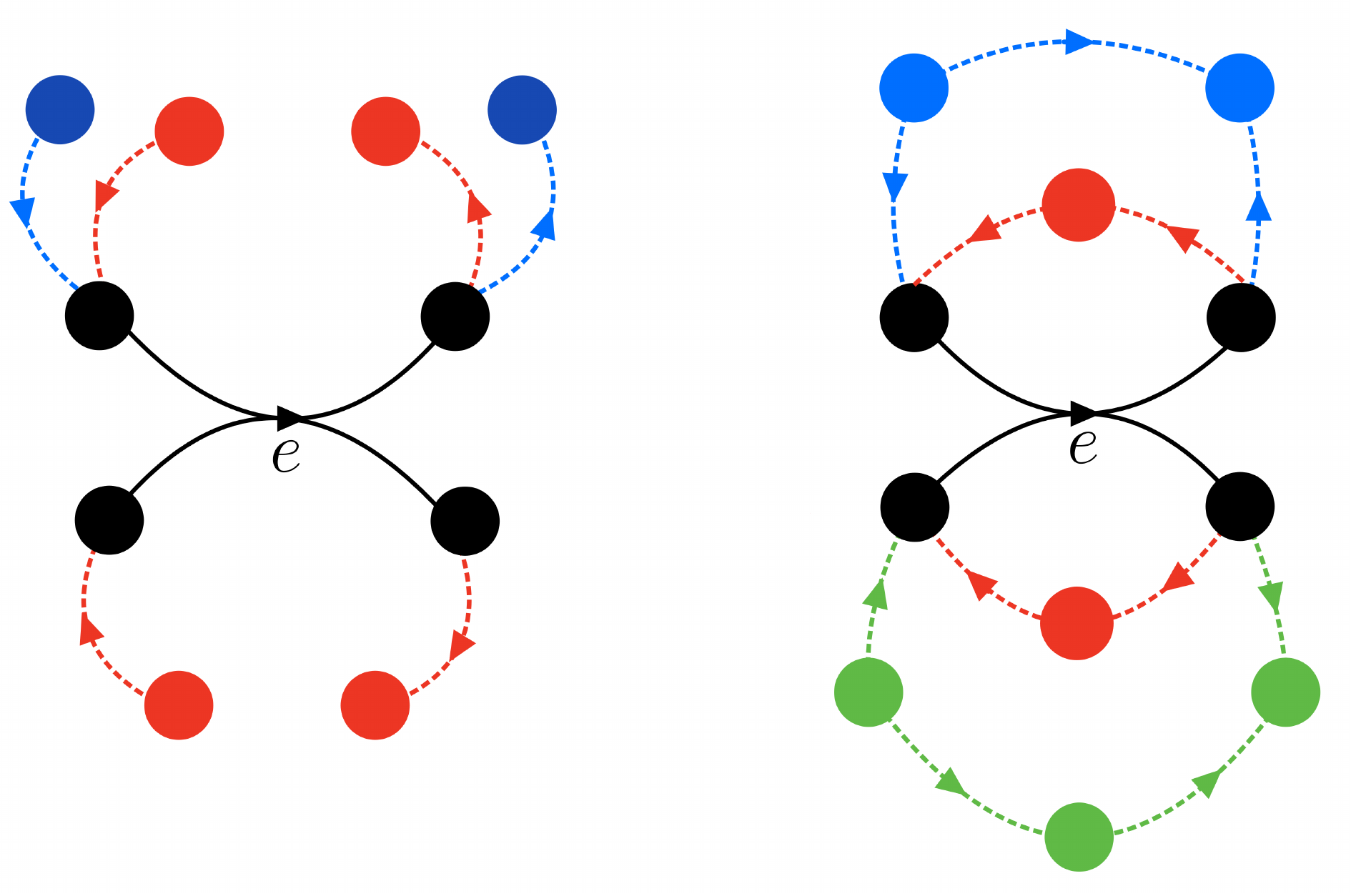}
	\caption{Illustration of the different connectivity patterns that affect $F(e) $ and $O(e)$. Well-connected hyperedges often play a key role in a network. Said hyperedges are identified by $F_{\rightarrow{}}(e)=|e_1|+|e_2|-\deg_{in}(e)-\deg_{out}(e)$, since it decreases monotonically with the number of incoming neighbors to the tail and outgoing neighbors from the head of $e$. Nevertheless, $F_{\rightarrow{}}(e)$ is not affected by the presence of arcs from the former to later. $O_{\rightarrow{}}(e)=m_0-m_2-2m_3$ captures this complementary information. On the right, we find directed triangles, which contribute to $ m_0 $ (black and red hyperedges), directed quadrangles to $ m_1 $ (black and blue hyperedges), and directed pentagons to $ m_2 $ (black and green hyperedges). In the figure on the left, the shortest path between any incoming and any outgoing neighbor, is 3. Such a connectivity pattern contributes to $ m_3 $.}
	\label{fig:FvsO}	 
\end{figure}

These two different curvature notions represent complementary tools for detecting local geometry and connectivity patterns in directed hypergraphs. Forman curvature  monotonically decreases with the number of incoming and outgoing neighbours of input and output nodes, resp., and it therefore detects hyperedges joining highly connected nodes. Ollivier curvature, on the other hand, is controlled by the overlap of the set of masses and holes (e.g. directed triangles) and by shortcuts between them (e.g. directed quadrangles and pentagons). We illustrate these principles in Figure \ref{fig:FvsO}. We want to evaluate the curvatures of the black hyperedge in the left and the right figure in various constellations. Without any of the colored hyperedges,  $F_{\rightarrow{}}(e)=|e_i| + |e_j|$, while $O_{\rightarrow{}}(e)=0$.   When the  red edges are present, we get $F_{\rightarrow{}}(e)=0$ in both the left and the right figure, whereas $O_{\rightarrow{}}(e)$ is negative in the left case, because there are no shortcuts,  but positive in the right case, when the inputs of the tail coincide with the outputs of the head. The presence of the blue edges on the right, however, makes a difference for $F_{\rightarrow{}}(e)$, but not for $O_{\rightarrow{}}(e)$, that is, the former, but not the latter distinguishes between those cases. In contrast, while $F_{\rightarrow{}}(e)$ does not distinguish between the presence of the blue and the  green edges in the right figure, $O_{\rightarrow{}}(e)$ sees the effect, as it is more negative in the presence of the green than in that of blue edges (the blue edges contribute to $m_1$, but the green ones to $m_2$).  

\subsection*{Metabolic networks}
\begin{figure}[t]                                   
\centering
\includegraphics[width=0.33\textwidth]{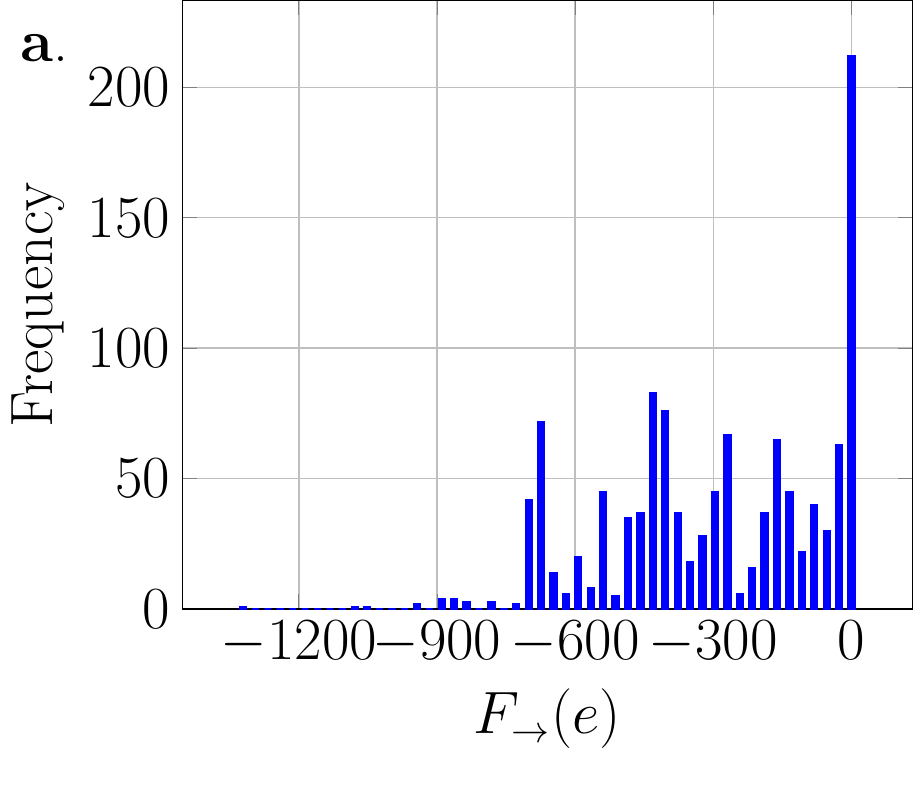}~
\includegraphics[width=0.33\textwidth]{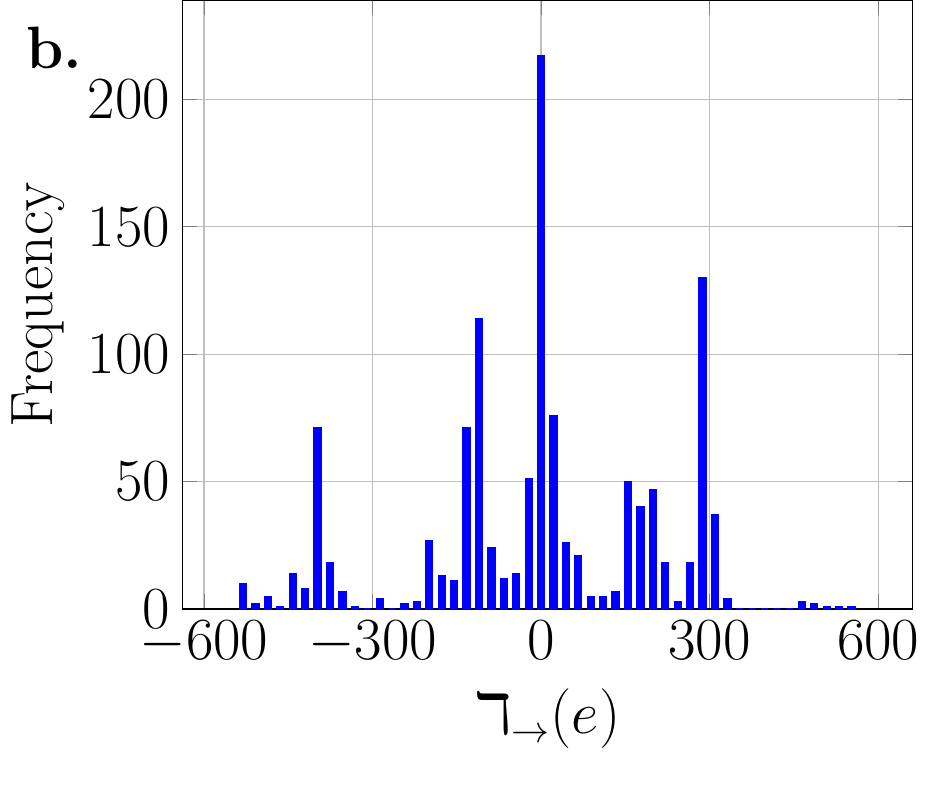}
\includegraphics[width=0.33\textwidth]{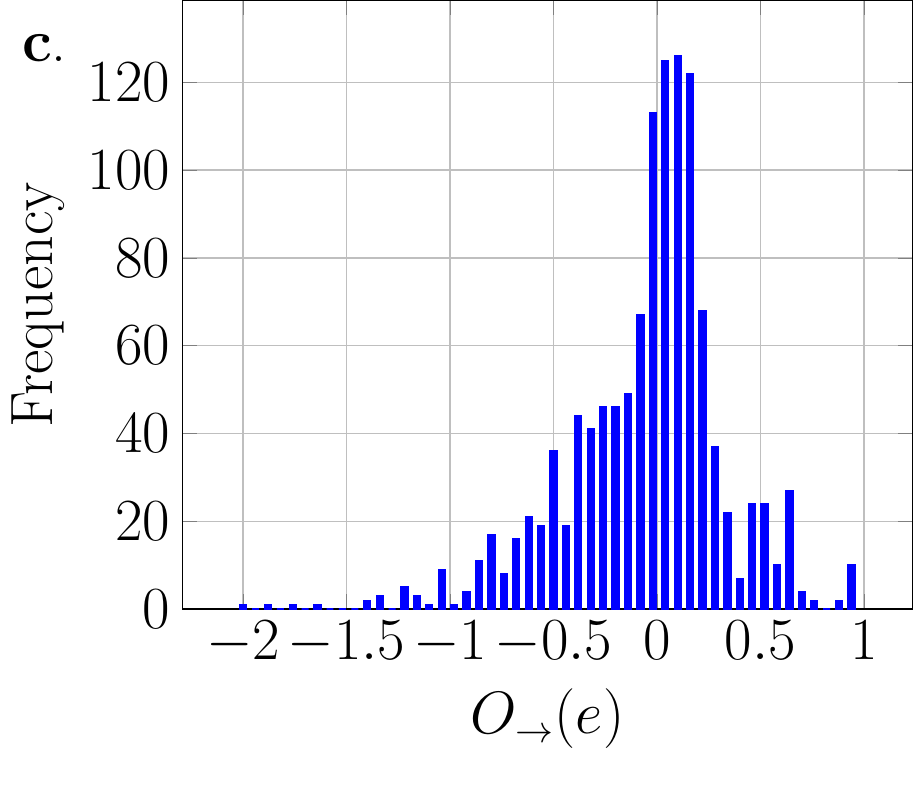}~
\caption{The distribution of (a) Forman-Ricci curvature, (b) degree difference, and (c) Ollivier-Ricci curvature in the metabolic network of \textit{Mycobacterium tuberculosis} H37Rv,  which is represented as a directed hypergraph with nodes as \textit{M. tuberculosis} metabolites and directed hyperedges as chemical reactions. The network has 743 nodes and 1195 hyperedge edges.}
\label{fig:FDDO_dist}
\end{figure}

\begin{figure}[t]                                   
\centering
\includegraphics[width=0.35\textwidth]{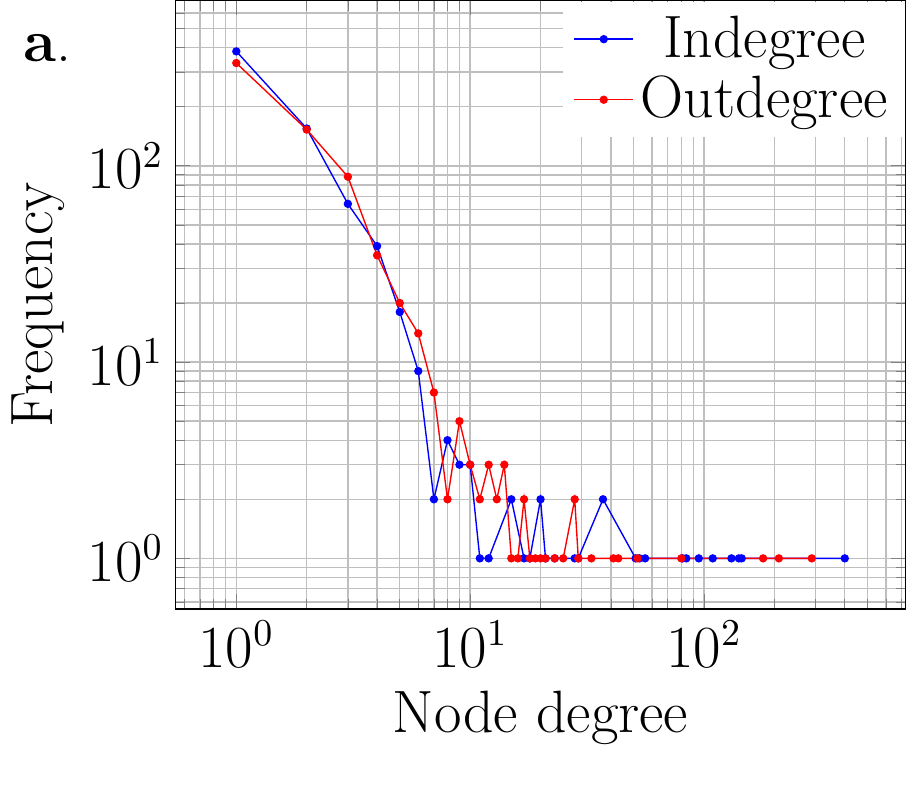}~
\includegraphics[width=0.325\textwidth]{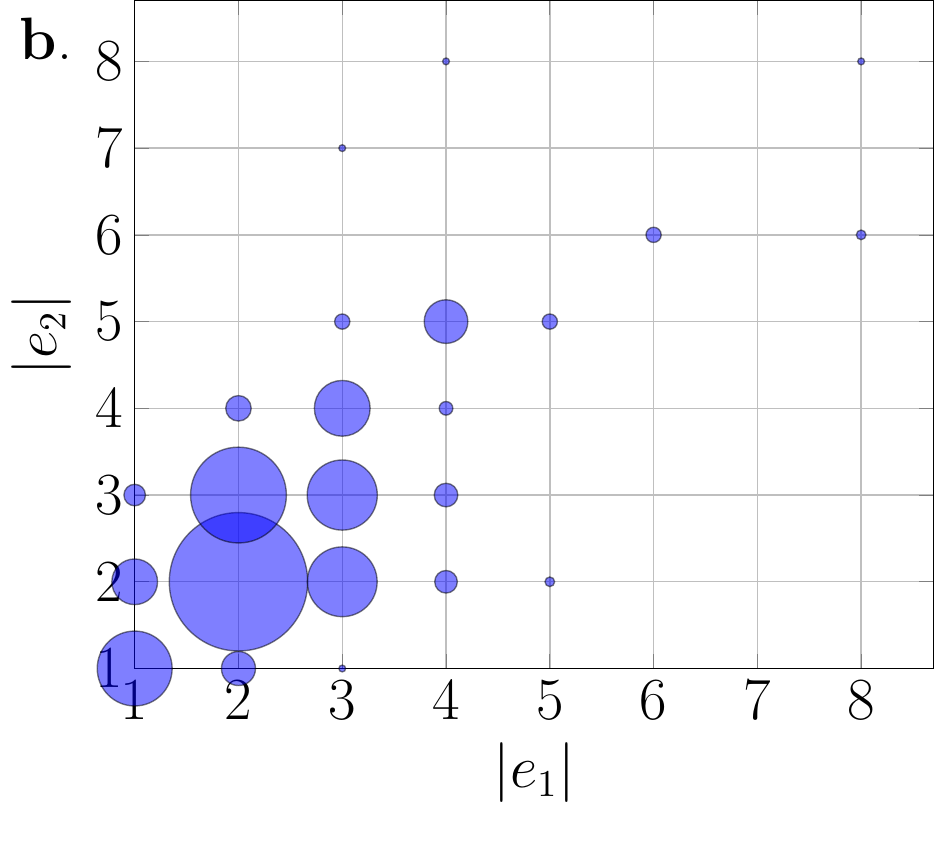}
\includegraphics[width=0.35\textwidth]{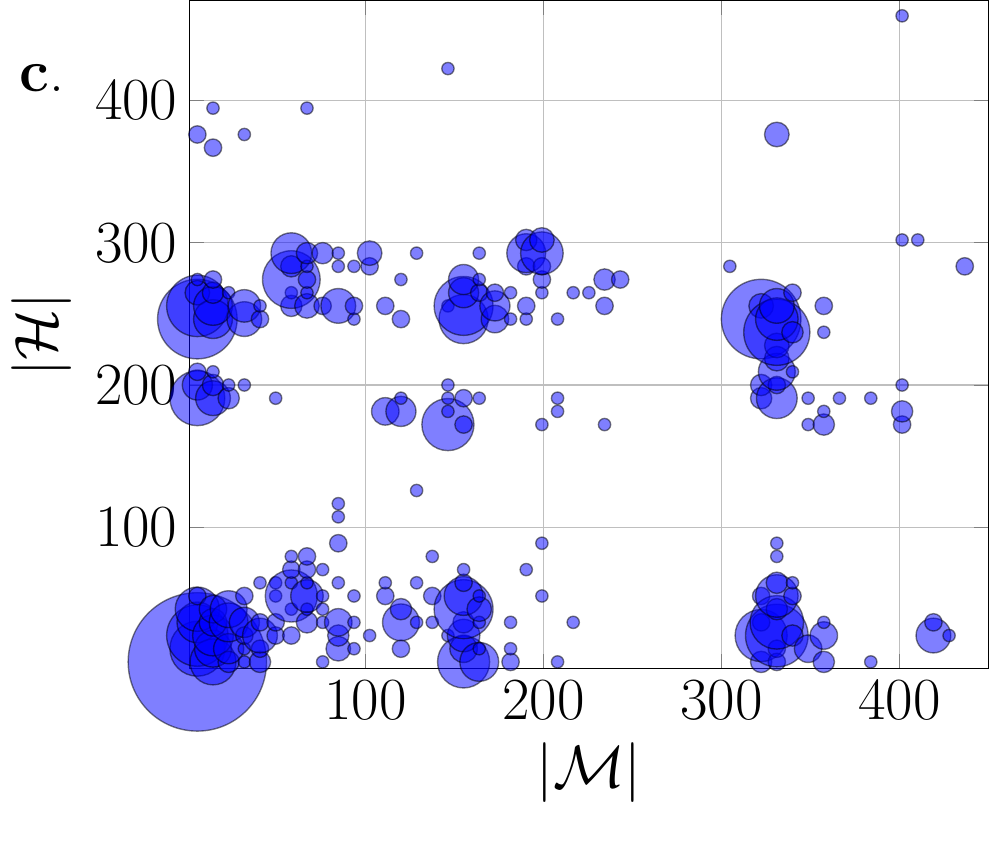}~
\includegraphics[width=0.35\textwidth]{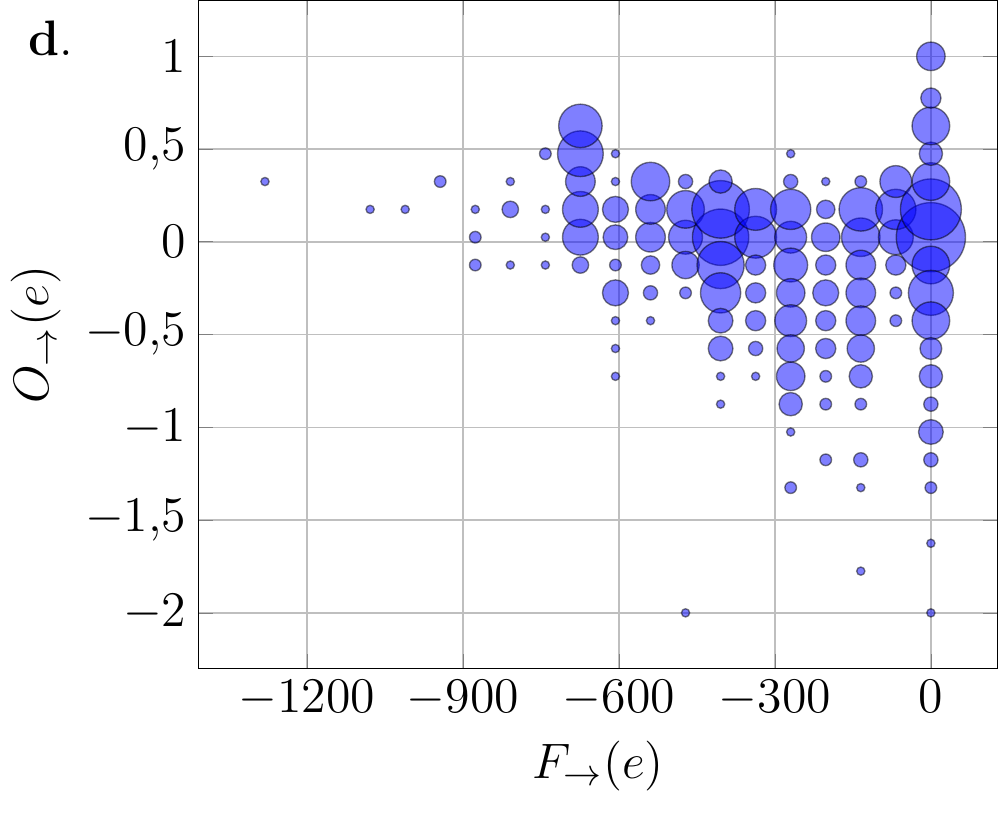}
\caption{The distribution of (a) node degree, (b) sizes of tails and heads, $(|e_1|,|e_2|)$, (c) number of masses and holes, $(\mathcal{M},\mathcal{H})$, and (d) $(\textrm{F}(e),\textrm{O}(e))$ values in the metabolic network of \textit{Mycobacterium tuberculosis} H37Rv,  which is represented as a directed hypergraph with nodes as \textit{M. tuberculosis} metabolites and directed hyperedges as chemical reactions. The network has 743 nodes and 1195 hyperedge edges.}
\label{fig:vs}
\end{figure}

Metabolic networks are evocative examples of directed hypergraphs, where  metabolites react with others to produce products. Both, reactants, $e_1$, and products, $e_2$, typically contain more than one substance ($|e_1|\geq 1$ or $|e_2|\geq 1$) and the reactions may not be reversible. This directed relationship between sets is therefore naturally modelled by a directed hyperedge ($e_1 \to e_2$). Since metabolic networks have been extensively studied, they present an ideal setting to illustrate how to use the hypergraph tools described here. For this, let us consider the metabolic network of \textit{Mycobacterium tuberculosis} H37Rv (version iNJ661) \cite{Jamshidi2007} modelled as a directed hypergraph. This network contains 939 reactions and 743 metabolites, of which 256 are reversible. Each reversible reaction ($e_1 \leftrightarrows e_2$) was divided into two, a forward reaction ($e_1 \rightarrow e_2$), and its reverse reaction ($e_1 \leftarrow e_2$). As a result, the network contains 1195 directed hyperedges. Most substrates of this metabolic network are consumed or produced by one reaction only. Also, a few are involved in more than half of the reactions ($\sim 50\%$ require \texttt{h}, \texttt{h2o}, \texttt{atp} or \texttt{nadhp}, and $\sim 57\%$ produce \texttt{h}, \texttt{pi}, \texttt{h2o}, \texttt{adp}, or \texttt{co2}), the distributions of indegree and outdegree in Figure \ref{fig:vs} a) summarize this behavior.\\
~
Suppose that we want to investigate whether starting materials that are produced in several different ways (large indegree in $e_1$) produce substances that also serve as starting materials for many reactions (large outdegree in $e_2$), that is, whether  targets are transformed into key precursors. There are  two aspects relevant for this question. First, we must find out if the network is assortative. Since it is a hypergraph, we use the degree difference and its distribution shows that this is mostly the case (see Figure \ref{fig:FDDO_dist} b)). Notoriously, the degree difference is $\sim 0$ for 217 ($\sim 18\%$) reactions. Second, we have to locate which of those 217 reactions involve metabolites of large degree. For that, instead of looking at the difference between out- and indegrees, we need to look at their sum and turn to  the distribution of $F_{\rightarrow{}}(e)$. Figure \ref{fig:FDDO_dist} a) shows that the dominant mode is represented by  curvature around zero. There are also secondary humps associated with more negative curvature values. Perhaps the most important reactions, however, are  those that have very low (negative) curvature values, but a  degree difference near zero. In fact,  the first reaction on the list is the fundamental reaction that creates the energy storage molecule adenosine triphosphate (ATP),  $e$: \texttt{adp+h+pi} $\to$ \texttt{atp+h+h2o}, with $F_{\rightarrow{}}(e)=-1347$ and $\daleth_{\rightarrow{}}(e) = 1$. Furthermore, the associated mass set $\mathcal{M}$ shows that there are 400 precursors for the substrates of this reaction, and, based on the set of holes $\mathcal{H}$, there are 464 derived metabolites. This pair of values correspond to the upper right blue mark in Figure \ref{fig:vs} c). Notice that this information is not given by node degree.  
With few exceptions, precursors and derivatives are at distances shorter than three, and mostly around zero, as shown by $O_{\rightarrow{}}(e)$ (see Figure \ref{fig:FDDO_dist} c)).  
For the reaction discussed here, $O_{\rightarrow{}}(e) = 0.35$ and it corresponds to the the left most blue mark of Figure \ref{fig:vs} d). The preceding already illustrates how a combination of the three measures that we have developed,  $F_{\rightarrow{}}(e), O_{\rightarrow{}}(e)$ and $\daleth_{\rightarrow{}}(e)$, can reveal the fundamental structural properties of specific reactions inside the metabolic network. Both evaluating the statistical distributions of these three quantities and comparing them for different  networks,  and analyzing those reactions that produce particularly prominent values for them in more detail should yield deeper insight into the structure of metabolic networks.

\section*{Acknowledgment}
W.L. would like to thank support from the German Academic Exchange Service (DAAD): Forschungsstipendien-Promotionen in Deutschland, 2017/2018 (Bewerbung 57299294).
A.S. would like to thank financial support from Max Planck Society Germany 
through the award of a Max Planck Partner Group in Mathematical Biology.

\bibliography{edges_arxiv}

\end{document}